\documentclass[%
 reprint,
superscriptaddress,
nofootinbib,
 amsmath,amssymb,
 aps,
]{revtex4-2}

\usepackage{graphicx}% Include figure files
\usepackage{dcolumn}% Align table columns on decimal point
\usepackage{bm}% bold math
\usepackage[colorlinks = true,
linkcolor = black,
urlcolor  = blue,
citecolor = black,
anchorcolor = black]{hyperref}
\usepackage{longtable}

\usepackage{chngcntr}
\begin{document}

\title{Electron Dynamics at High-Energy Densities in Nickel from Non-linear Resonant X-ray Absorption Spectra}

%\thanks{A footnote to the article title}%
%%%%%%%%%%%%%%%%%%%%%%%%%%%%%%%%%%%%%%%%%%%%%%%%%%%%%%%%%%%%%%%%%%%%%%%%%%%%%%%%%%%%%%%%%%%%%%%%%%%%%%%%%%%%%%%%%%%%%%%%%%%%%%%%%%%
\author{Robin Y. Engel}
    \affiliation{Deutsches Elektronen-Synchrotron DESY, Notkestr. 85, 22607 Hamburg, Germany}%
    \affiliation{Department of Physics, Universität Hamburg, Luruper Chaussee 149, 22761 Hamburg, Germany}
\author{Oliver Alexander}
    \affiliation{Imperial College London, Department of Physics, Exhibition Rd, London SW7 2BX, United Kingdom}
\author{Kaan Atak}
    \affiliation{Deutsches Elektronen-Synchrotron DESY, Notkestr. 85, 22607 Hamburg, Germany}%
\author{Uwe Bovensiepen}
    \affiliation{Faculty of Physics and Center for Nanointegration Duisburg-Essen (CENIDE), University of Duisburg-Essen, Lotharstr. 1, 47057 Duisburg, Germany}
    \affiliation{Institute for Solid State Physics, University of Tokyo, Kashiwa, Chiba, 277-8581, Japan}
\author{Jens Buck}
    \affiliation{Deutsches Elektronen-Synchrotron DESY, Notkestr. 85, 22607 Hamburg, Germany}%
    \affiliation{Christian-Albrechts-Universität zu Kiel, Institut für Experimentelle und Angewandte Physik, Leibnizstraße 11-19, 24118 Kiel, Germany}
\author{Robert Carley}
    \affiliation{European XFEL, Holzkoppel 4, 22869 Schenefeld, Germany}%
\author{Michele Cascella}
    \affiliation{MAX IV Laboratory, Lund University, PO Box 118, SE-221 00 Lund, Sweden}%
\author{Valentin Chardonnet}
    \affiliation{Sorbonne Université, CNRS, Laboratoire de Chimie Physique-Matière et Rayonnement, 4 Pl. Jussieu Barre 43-44, 75005 Paris, France}
\author{Gheorghe Sorin Chiuzbaian}
    \affiliation{Sorbonne Université, CNRS, Laboratoire de Chimie Physique-Matière et Rayonnement, 4 Pl. Jussieu Barre 43-44, 75005 Paris, France}
\author{Christian David}
    \affiliation{Paul Scherrer Institut, Forschungsstrasse 111, 5232 Villigen, Switzerland}%
\author{Florian Döring}
    \affiliation{Paul Scherrer Institut, Forschungsstrasse 111, 5232 Villigen, Switzerland}%
\author{Andrea Eschenlohr}
    \affiliation{Faculty of Physics and Center for Nanointegration Duisburg-Essen (CENIDE), University of Duisburg-Essen, Lotharstr. 1, 47057 Duisburg, Germany}
\author{Natalia Gerasimova}
    \affiliation{European XFEL, Holzkoppel 4, 22869 Schenefeld, Germany}%
\author{Frank de Groot}
    \affiliation{Utrecht University, Debye Institute for Nanomaterials Science, Inorganic Chemistry and Catalysis, Princetonplein 1, Universiteitsweg 99, 3584 CC Utrecht, Netherlands}
\author{Lo\"ic Le Guyader}
    \affiliation{European XFEL, Holzkoppel 4, 22869 Schenefeld, Germany}%
\author{Oliver S. Humphries}
    \affiliation{European XFEL, Holzkoppel 4, 22869 Schenefeld, Germany}%
\author{Manuel Izquierdo}
    \affiliation{European XFEL, Holzkoppel 4, 22869 Schenefeld, Germany}%
\author{Emmanuelle Jal}
    \affiliation{Sorbonne Université, CNRS, Laboratoire de Chimie Physique-Matière et Rayonnement, 4 Pl. Jussieu Barre 43-44, 75005 Paris, France}
\author{Adam Kubec}
    \affiliation{Paul Scherrer Institut, Forschungsstrasse 111, 5232 Villigen, Switzerland}%
\author{Tim Laarmann}
    \affiliation{Deutsches Elektronen-Synchrotron DESY, Notkestr. 85, 22607 Hamburg, Germany}%
    \affiliation{The Hamburg Centre for Ultrafast Imaging CUI, Luruper Chaussee 149, 22761 Hamburg, Germany}
\author{Charles-Henri Lambert}
    \affiliation{Department of Materials, ETH Zurich, 8093 Zurich, Switzerland}
\author{Jan Lüning}
    \affiliation{Helmholtz-Zentrum Berlin für Materialien und Energie GmbH, Hahn-Meitner-Platz 1, 14109 Berlin, Germany}
\author{Jonathan P. Marangos}
    \affiliation{Imperial College London, Department of Physics, Exhibition Rd, London SW7 2BX, United Kingdom}
\author{Laurent Mercadier}
    \affiliation{European XFEL, Holzkoppel 4, 22869 Schenefeld, Germany}%
\author{Giuseppe Mercurio}
    \affiliation{European XFEL, Holzkoppel 4, 22869 Schenefeld, Germany}%
\author{Piter S. Miedema}
    \affiliation{Deutsches Elektronen-Synchrotron DESY, Notkestr. 85, 22607 Hamburg, Germany}%
\author{Katharina Ollefs}
    \affiliation{Faculty of Physics and Center for Nanointegration Duisburg-Essen (CENIDE), University of Duisburg-Essen, Lotharstr. 1, 47057 Duisburg, Germany}
 \author{Bastian Pfau}
    \affiliation{Max Born Institute for Nonlinear Optics and Short Pulse Spectroscopy, Max-Born-Str. 2A, 12489 Berlin, Germany}
\author{Benedikt Rösner}%
    \affiliation{Paul Scherrer Institut, Forschungsstrasse 111, 5232 Villigen, Switzerland}%    
\author{Kai Rossnagel}
    \affiliation{Deutsches Elektronen-Synchrotron DESY, Notkestr. 85, 22607 Hamburg, Germany}%
    \affiliation{Christian-Albrechts-Universität zu Kiel, Institut für Experimentelle und Angewandte Physik, Leibnizstraße 11-19, 24118 Kiel, Germany}
\author{Nico Rothenbach}
    \affiliation{Faculty of Physics and Center for Nanointegration Duisburg-Essen (CENIDE), University of Duisburg-Essen, Lotharstr. 1, 47057 Duisburg, Germany}
\author{Andreas Scherz}
    \affiliation{European XFEL, Holzkoppel 4, 22869 Schenefeld, Germany}%
\author{Justine Schlappa}
    \affiliation{European XFEL, Holzkoppel 4, 22869 Schenefeld, Germany}%
\author{Markus Scholz}
    \affiliation{Deutsches Elektronen-Synchrotron DESY, Notkestr. 85, 22607 Hamburg, Germany}%
    \affiliation{European XFEL, Holzkoppel 4, 22869 Schenefeld, Germany}%
\author{Jan O. Schunck}
    \affiliation{Deutsches Elektronen-Synchrotron DESY, Notkestr. 85, 22607 Hamburg, Germany}%
    \affiliation{Department of Physics, Universität Hamburg, Luruper Chaussee 149, 22761 Hamburg, Germany}
\author{Kiana Setoodehnia}
    \affiliation{European XFEL, Holzkoppel 4, 22869 Schenefeld, Germany}
\author{Christian Stamm}
    \affiliation{Department of Materials, ETH Zurich, 8093 Zurich, Switzerland}
    \affiliation{Institute for Electric Power Systems, University of Applied Sciences and Arts Northwestern Switzerland, 5210 Windisch, Switzerland}
\author{Simone Techert}
    \affiliation{Deutsches Elektronen-Synchrotron DESY, Notkestr. 85, 22607 Hamburg, Germany}%
    \affiliation{Institute for X-ray Physics, Goettingen University, Friedrich Hund Platz 1, 37077 Goettingen, Germany}
\author{Sam M. Vinko}
    \affiliation{Department of Physics, Clarendon Laboratory, University of Oxford, Parks Road, Oxford OX1 3PU, United Kingdom}
    \affiliation{Central Laser Facility, STFC Rutherford Appleton Laboratory, Didcot OX11 0QX, United Kingdom}
\author{Heiko Wende}
    \affiliation{Faculty of Physics and Center for Nanointegration Duisburg-Essen (CENIDE), University of Duisburg-Essen, Lotharstr. 1, 47057 Duisburg, Germany}
\author{Alexander A. Yaroslavtsev}
    \affiliation{Uppsala University, Department of Physics and Astronomy, Regementsvägen 1 Uppsala, Sweden}
\author{Zhong Yin}
    \affiliation{International Center for Synchrotron Radiation Innovation Smart, Tohoku University, 2-1-1 Katahira, Aoba-ku, Sendai, Miyagi 980-8577, Japan}%
    \affiliation{ETH Zürich, Laboratorium für Physikalische Chemie, Vladimir-Prelog-Weg 1-5, 8093 Zürich, Switzerland}
\author{Martin Beye}
    \affiliation{Deutsches Elektronen-Synchrotron DESY, Notkestr. 85, 22607 Hamburg, Germany}%
    \affiliation{Department of Physics, Universität Hamburg, Luruper Chaussee 149, 22761 Hamburg, Germany}
    \email{martin.beye@desy.de}
\date{\today}% It is always \today, today,
             %  but any date may be explicitly specified
%%%%%%%%%%%%%%%%%%%%%%%%%%%%%%%%%%%%%%%%%%%%%%%%%%%%%%%%%%%%%%%%%%%%%%%%%%%%%%%%%%%%%%%%%%%%%%%%%%%%%%%%%%%%%%%%%%%%%%%%%%%%%%%%%%%

\begin{abstract}
The pulse intensity from X-ray free-electron lasers (FELs) can create extreme excitation densities in solids, entering the regime of non-linear X-ray-matter interactions.
We show $L_3$-edge absorption spectra of metallic nickel thin films with fluences 
entering a regime where several X-ray photons are incident per absorption cross-section.
Main features of the observed non-linear spectral changes are described with a predictive rate model for electron population dynamics during the pulse, utilizing a fixed density of states and tabulated ground-state properties.%, while others call for more detailed simulations. 
\end{abstract}

\maketitle

The modern understanding of complex solid materials relies on appropriate approximations to the unabridged quantum mechanical description of the full, correlated many-body problem. % Gute Quelle für die Aussage?
To assess the predictive power of theoretical models and the selected approximations, detailed experimental studies far away from known territory are especially insightful. %required.
Absorbing the high power densities available from an X-ray free-electron laser (FEL) in a solid metal generates a very unusual state of warm dense matter far from equilibrium: Individual electronic excitations reach up to hundreds of eV and excitation levels average out to many eV per atom  
\cite{Zasrtrau2008aluminum, Vinko2012solid_plasma, Cho2012solidPlasma, Humphries2020WDNi, GarciaSaiz2008WDM_RIXS, Vinko2015DensePlasmaXFELs, Bailey2015WDironOpacity, Hollebon2019WDalumAbInitio, Preston2017WDmagnesium}.  % Studies on X-ray WDM but not NL-Absorption,
As the absorption of an intense X-ray pulse depends on the changes it drives in the electronic system \cite{bob2009turning, Recoules2009alum_XANES, cicco2014interplay, rackstraw2015saturable, principi2016free, Chen2018self_induced_transparency, yoneda2014saturable, Chen2018self_induced_transparency}, % Studies on solids with NL-absorption.
a single-pulse non-linear absorption measurement can be used to investigate its evolution on the timescale of the pulse duration.

We present fluence-dependent X-ray absorption spectra recorded with monochromatic X-rays on metallic nickel thin films around the nickel 2$p_{3/2}$ ($L_3$) edge, revealing a changing valence electron system around the Fermi level as a consequence of the high excitation densities from fluences up to 60 J/cm$^2$ (corresponding to 2$\times10^{15}$ W/cm$^{2}$).

The electronic processes that ensue after the absorption of photons at core levels trigger a complex dynamical process that is challenging to treat in purely \textit{ab-initio} simulations \cite{Chen2020abi_initio_transabs,  Mo2020firstPriciplesThomson, williams2020collisional, medvedev2018various, lipp2022density}. % Lipp writes: "In case of X-ray irradiation, additional obstacle is high energy of X-ray excited electrons (up to keV). Such range of electronic energies is challenging to be treated with a fully ab-initio method27."
Here, we take an alternative approach and develop a simple rate equation model that provides an intuitive understanding of the relevant processes \cite{engel2022model}.
The resulting picture of the evolution of electronic populations within a fixed ground-state density of states successfully describes the largest part of the non-linear changes in the spectra. This corroborates the dominant impact of electron redistribution from the strong non-equilibrium state towards a thermalized electronic system. Some of the observed changes, especially in the close vicinity of the resonance, deviate from the predictions of the rate model and call for more evolved theories. Here, our work provides a benchmark to identify observations of advanced physical processes and effects. While this letter discusses the experiment and resulting insights, we lay out the framework of the model in detail in a separate publication \cite{engel2022model}.

% Relating to NL-spectroscopy
Additionally, our straightforward picture of intense core-resonant X-ray pulse interaction with the valence system of a 3$d$ metal lays a solid knowledge-based foundation for the planning and interpretation of non-linear X-ray spectroscopy experiments at FELs; in particular, the relevance of electronic scattering processes observed here is expected to affect methods relying on stimulated emission from core excitations and X-ray or X-ray/optical wave-mixing \cite{mukamel2005multiple, Glover2012xray, beye2013stimulated, weninger2013stimulated, Shwartz2014XSHG, Tamasaku2014TPA_competing, Bencivenga2015FWMTG, schreck2015implications, Lam2018HXSHGinterfacial, Tamasaku2018NL_TPA, higley2019femtosecond, rottke2021probing, Rouxel2021HXTG, Bencivenga2021FWMtwo_color, Wirok_Stoletow2022TPA_Ge, higley2022stimulated}. % X-ray inner-shell laser not mentioned because it is not an analysis method and not on solids

%\section{Main}

\begin{figure}[ht]
\centering
\includegraphics[width=\linewidth]{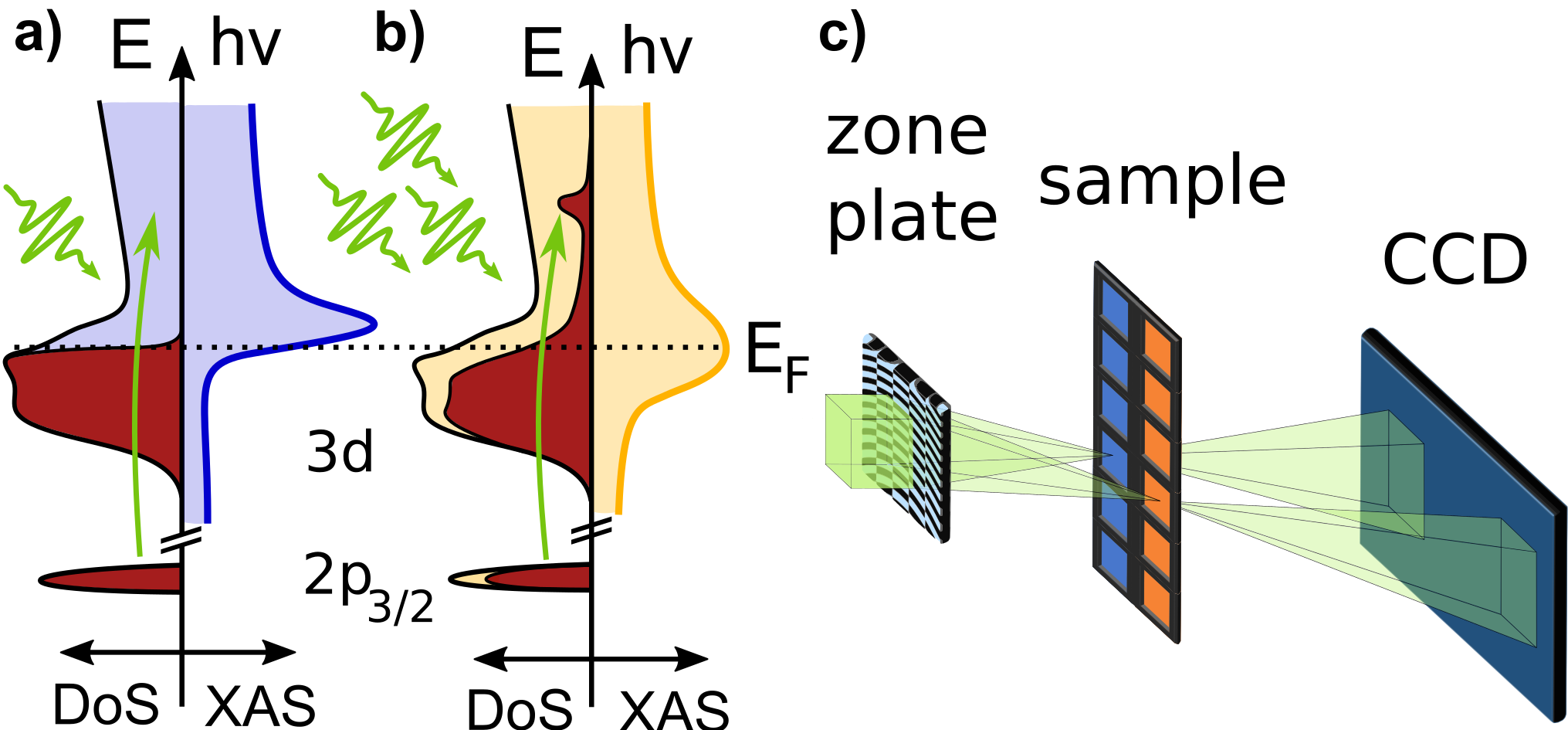}
\caption{
	\textbf{(a\&b) Sketch of absorption at different fluences.} The unoccupied states determine the XAS spectrum as they are probed by core-resonant photons.
	\textbf{(a)} In the low-fluence case  (blue unoccupied states and resulting spectrum), the electronic system mostly remains in the ground state. \textbf{(b)} In the high-fluence case (yellow unoccupied states and spectrum), later parts of the X-ray pulse probe a hot electronic system and experience spectral bleaching at the probed photon energy.\\
	\textbf{Setup for non-linear XAS (c)} The split-beam-normalization scheme uses a special zone plate \cite{Doring2020zoneplates}, which generates two adjacent beam foci for transmission through the sample and a reference membrane before the beams impinge on the detector.
	}
\label{fig: setup}
\end{figure}

%\section{\label{sec: experiment} Experiment}
X-ray absorption spectra of the nickel 2$p_{3/2}$ ($L_3$) edge were recorded at the Spectroscopy and Coherent Scattering Instrument (SCS) of the European XFEL \cite{Tschentscher2017xfel}.

The XAS spectra were measured by continuously scanning the SASE3 monochromator \cite{Gerasimova2022mono} (synchronized with the undulator gap) back-and-forth many times in the range 846-856\,eV. The photon bandwidth was about 420 meV and the FEL pulse duration on the sample was about 30 fs FWHM. The overall beam intensity was controlled using a gas attenuator filled with nitrogen and monitored using an X-ray gas-monitor (XGM) downstream of the monochromator \cite{Grunert2019xfelDiag,Maltezopoulos2019operation}. 

For X-ray absorption measurements at FELs based on Self-Amplified Spontaneous Emission (SASE), beam-splitting schemes can deliver optimal normalization of SASE-fluctuations \cite{engel2021shot, schlotter2020balanced, guyader2022shot_noise}. Here, a focusing and beam-splitting zone plate also creates the required tight focusing to achieve extreme fluences.
Figure \ref{fig: setup} shows the schematic experimental layout.

The zone plate combines an off-axis Fresnel structure for focusing and a line grating for beam-splitting in a single optical element \cite{Doring2020zoneplates}.
It thus produces two $\mu$m-sized, identical foci in the sample plane, 1.9\,mm apart, originating from the first-order diffraction of the zone plate, as well as the positive and negative first orders of the line grating.
The sample has a square support of 25\,mm size, containing  %20 $\times$ 20 square
Si$_3$N$_4$ membrane windows (orange in Figure \ref{fig: setup}) of 0.5\,mm size and 200\,nm thickness with a distance of 2\,mm between adjacent windows. 
Every second pair of rows (blue in Figure \ref{fig: setup}) was additionally coated with a 20\,$\mathrm{nm}$ sample layer of polycrystalline metallic Ni by sputter deposition, on top of a 2\,$\mathrm{nm}$ bonding layer of Ta; a 2\,$\mathrm{nm}$ Pt capping layer prevents oxidation during sample-handling.

The sample frame was positioned such that one zone plate focus %(the \textit{signal}) 
impinged on a nickel-coated membrane, while the other %(the \textit{reference})
hit a bare silicon-nitride membrane. Thus, the difference in transmission of both beams can be attributed solely to the nickel film.

The detector was a fast readout-speed charge-coupled device (FastCCD) with high dynamic range, enabling 10\,Hz read-out and increasing the fluence range available to the experiment \cite{denes2009ccd, januschek2016performance, klakov2019ccd}.
Due to the unstable detector temperature, significant retroactive calibration of the detector was necessary (see supplement).
To prevent detector saturation, an additional aluminum filter of about 13\,$\mathrm{\mu m}$ thickness was used between sample and detector for measurements with the unattenuated beam.

During these high-intensity measurements, sample and reference films were locally damaged by intense individual FEL shots. Thus, the FEL was operated in single-shot mode at 10\,Hz repetition rate, and the sample was scanned through the beam continuously at 0.5\,$\mathrm{mm \cdot s^{-1}}$, resulting in 10 shots per membrane window. 

The shot craters in the reference membranes were later analyzed with scanning electron microscopy (SEM) to determine the effective focal size at specific photon energies. The resulting spot sizes were used to calibrate ray-tracing calculations which delivered the photon-energy-dependent spot size, ranging from 0.4\,$\mathrm{\mu m^2}$ to about 3\,$\mathrm{\mu m^2}$ (see supplement for details on the spot size determination).

\begin{figure}
    \centering
    \includegraphics[width = \linewidth]{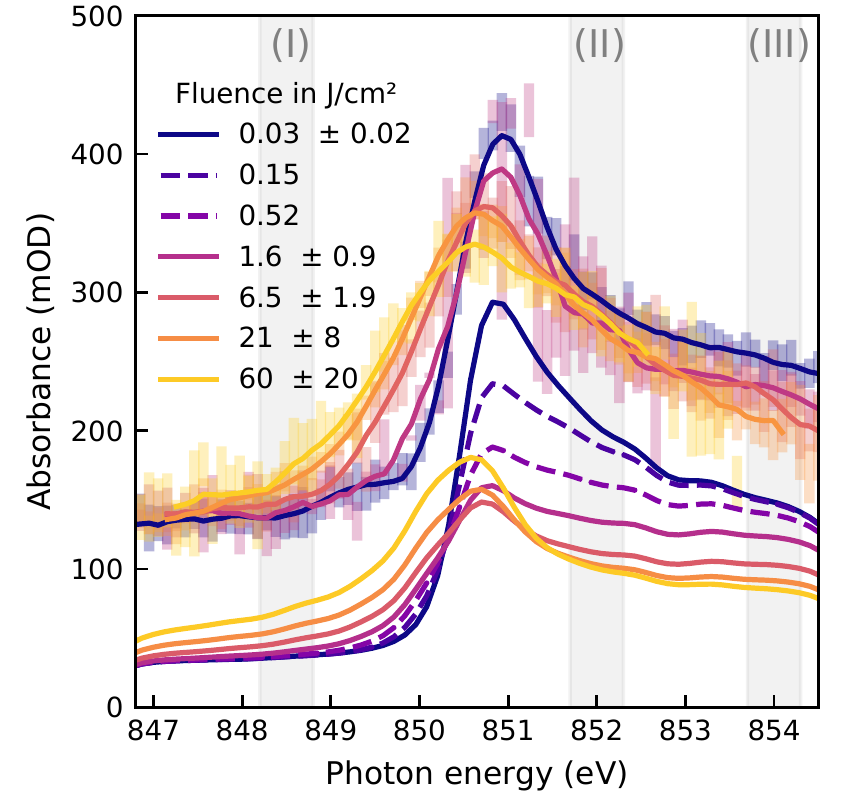}
    \caption{\textbf{Fluence-dependent Ni\,$L_3$-edge spectra, measured (top) and simulated (bottom).} The fluence of events contributing to each spectrum is given in the legend in terms of mean and standard deviation. Dashed simulated spectra do not have a corresponding measurement.
    The regions of interest from which absorbance changes shown in panels b), d), and e) of Figure \ref{fig: effect comparison} were quantified are shaded and labeled (I), (II), and (III), respectively.
    The error bars are shown for the measured spectra and represent the 95\% confidence intervals for each bin of 102\,meV width; solid lines of the measured spectra are smoothed using a Savitzky-Golay filter using windows of 21 bins and 4th-order polynomials.  The experimental spectra are vertically offset by 100\,mOD.
    %Du könntest noch die Labels b, d-e an die grauen Boxen machen und auf Figure 5 verweisen.
    }
    \label{fig: spectra}
\end{figure}

Figure \ref{fig: spectra} shows the spectra for the nickel $L_3$-edge next to simulated spectra for increasing X-ray fluence over more than 3 orders or magnitude, from 0.03 to 60 J/cm$^2$. Each measured point represents an average of several FEL shots, sorted by X-ray fluence and photon energy. The varying statistical uncertainty is a result of the pulse intensity fluctuations of monochromatized SASE radiation \cite{saldin2006statistical} in combination with photon energy-dependent spot sizes (see supplement for details on the shot sorting).

\begin{figure}
    \centering
    \includegraphics[width = \linewidth]{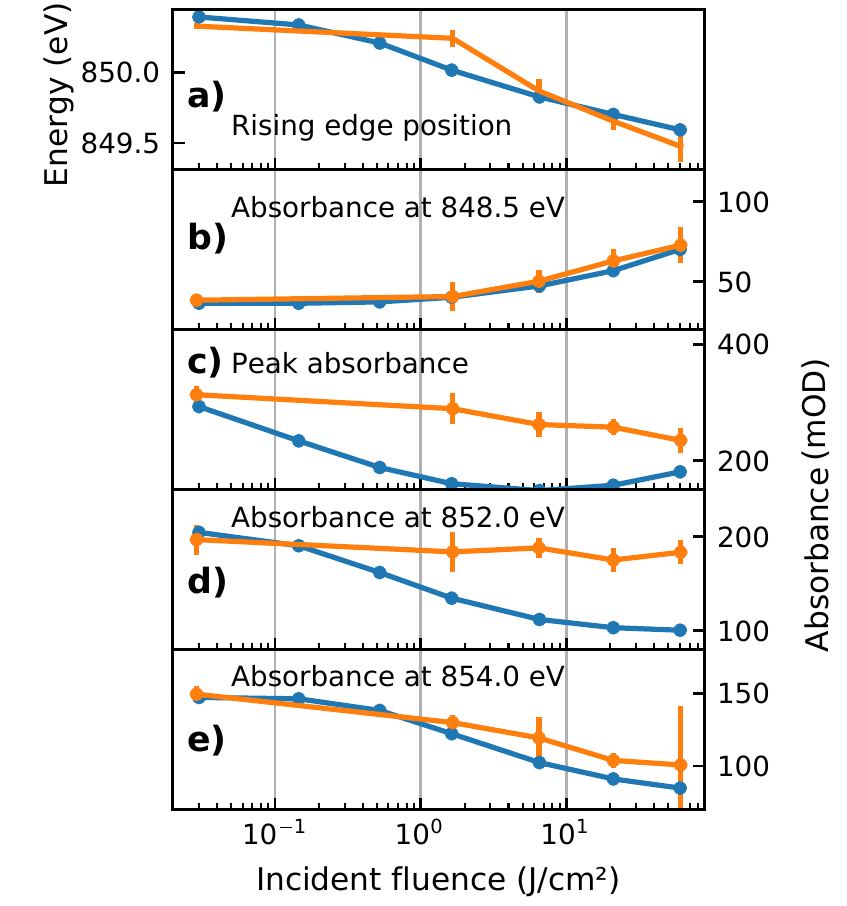}
    \caption{\textbf{Comparison of spectral effects} between simulation (blue lines) and experiment (orange lines with error bars). The shift of absorption edge in panel a) represents the photon energy at which the half-maximum of the absorption peak is reached. The absorbance changes in panels b), d) and e) are integrated from the gray shaded regions in Figure \ref{fig: spectra}, while panel c) shows the global maximum of the spectrum.}
    \label{fig: effect comparison}
\end{figure}

% Define observables
We observe four main fluence-dependent effects, which we quantify and compare to the simulated results in Figure \ref{fig: effect comparison}: a) a red-shift of the absorption peak of up to 0.9\,$\pm$0.1\,eV in the rising flank; b) an increase of the pre-edge absorbance, as the rising edge of the absorption peak shifts and broadens; c) a reduced peak absorbance and d), e) a reduced post-edge absorbance.
The integration regions from which the effects b), d) and e) are derived, are highlighted in Figure \ref{fig: spectra} as (I), (II) and (III), respectively.
The shift of the absorption edge is quantified by the photon energy at which the absorbance reaches half of the peak value; its uncertainty is propagated from the statistical uncertainty of the absorption peak measurement. 

Before we analyze these observations in detail, let us quickly paraphrase our modeling approach \cite{engel2022model}:
In contrast to earlier rate models \cite{hatada2014modeling, cicco2014interplay}, we describe the evolution of the electronic system with an energy-resolved population of the valence band. 
Tracking the full non-thermal population history proved crucial to describe the non-linear absorption changes near and around the Fermi level.
As coupling between electrons and phonons in metals is typically not yet important on the timescale of 30\,fs \cite{Anisimov1973Two_Temp, chen_two_temp, Hartley2019disordering} and we do not account for collective electron correlation effects, we test the approximation that the Density of States (DoS) remains unchanged within the pulse duration.

Transition rates between electronic states are determined by scaling  ground state rates with the populations of initial and final states.
The relevant process rates are compiled into differentials of electronic populations and photon density in space and time and implemented in a finite-element simulation to derive the electron population history and ultimately the X-ray transmission of a three-dimensional sample.

The model implements the processes of resonant absorption from the 2$p_{3/2}$ core level and non-resonant absorption from other (mostly valence) electrons. Stimulated emission is described as a time-inverted resonant absorption process. 
Electronic thermalization is modeled with a bulk timescale $\tau_{\mathrm{th}}$ (essentially quantifying electron-electron scattering) that moves the non-thermal valence electron distribution towards a target Fermi-Dirac distribution that corresponds to the momentary internal energy and population of the valence band. Finally, scattering cascades initiated by fast Auger-electrons and photo-electrons from non-resonant absorption are parameterized by another scattering time $\tau_{\mathrm{scatt}}$. 

With this simple description of the underlying processes, we provide a microscopic picture of the electronic system and its interaction with resonant X-rays as a complementary approach to more complex calculations \cite{medvedev2018various,lipp2022density}. %and thus contribute to the development of non-linear X-ray spectroscopy.

% Discuss agreement
Solely considering the population dynamics of the electronic system, the simulation already achieves good agreement with the experimental data across more than three orders of magnitude in fluence.
% Only 2 free parameters
This is particularly remarkable since nearly all input parameters are experimental parameters or well-known ground-state properties of the material, such as density, electronic configuration, and ground-state spectrum.
Only the valence thermalization time $\tau_{\mathrm{th}}$ and electron scattering time $\tau_{scatt}$ were varied to achieve the best match to the experimental results. The found value $\tau_{\mathrm{th}}=6$\,fs compares well to recent estimates for excitations on this energy scale \cite{higley2019femtosecond, higley2022stimulated,mueller2013relaxation, chang2021electron}.
% Scattering

The time constant $\tau_\mathrm{{scatt}}=1.5$\,fs characterizes the secondary electron scattering cascade which transfers energy (and population) from fast electrons to (unoccupied) valence states. The constant summarizes many individual electron scattering events and compares to the tabulated time between individual collisions in ground-state nickel of roughly 100 attoseconds \cite{powell2020NIST}.

\begin{figure}[h]
\centering
\includegraphics[width=\linewidth]{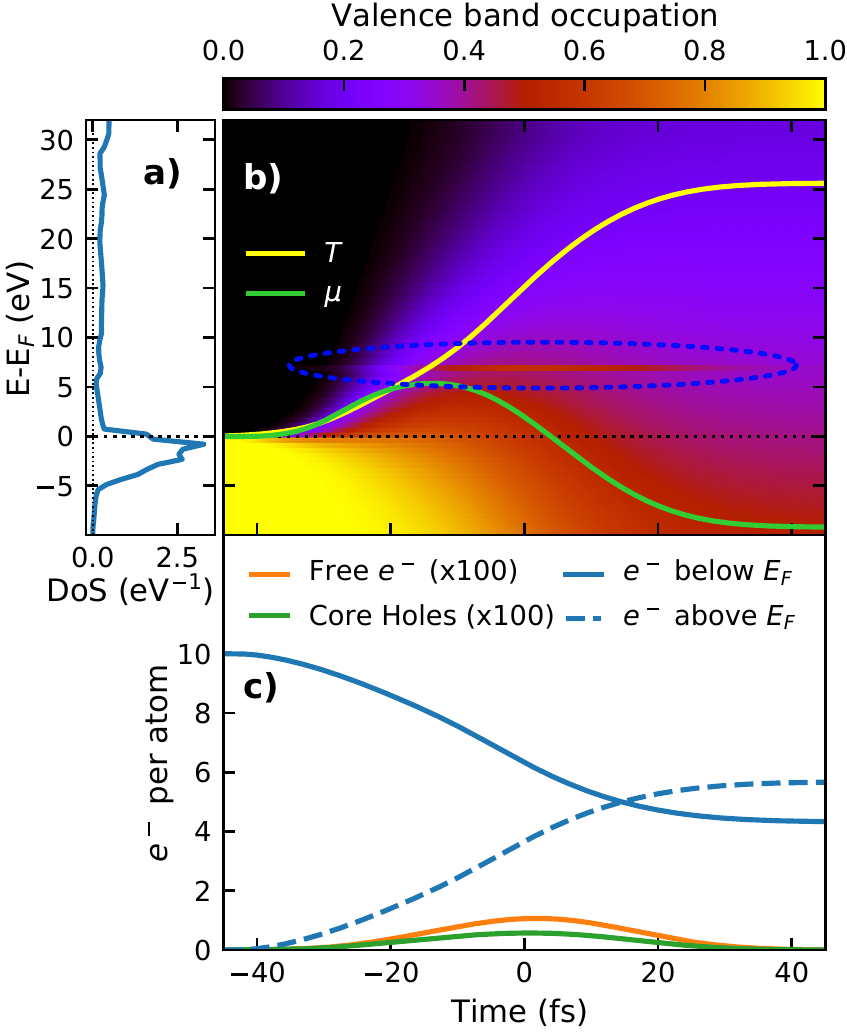}
\caption{
	\textbf{Evolution of electronic populations (simulation)} in a single voxel at the sample surface for a pulse of 858.3\,eV, with a pulse energy of 30\,$\mathrm{J/cm^2}$. Panel a) shows the total DoS used as an input for the simulation. Panel b) shows the energy-resolved (relative to the Fermi energy) occupation of the valence band over time. The population (in electrons/atom/eV) is the product of the DoS and the occupation.	The thermalized valence occupation lags a few femtoseconds behind the current chemical potential $\mu$; the temperature $T$ of the valence system rises rapidly, ultimately reaching up to 25\,eV. The bleaching of valence states (highlighted with a blue dotted ellipse) is visible as a high non-thermal population at the resonant photon energy around 7\,eV above the Fermi level.
	Panel c) shows the number of core holes and free electrons over time, as well as the number of electrons in the valence system below and above the Fermi energy.}
\label{fig: dos development}
\end{figure}

% Example of the electron dynamics
Figure \ref{fig: dos development} shows an example of a simulated valence band population history, specifically from the uppermost 4\,\AA{} thick layer of the sample, excited with a Gaussian pulse profile centered around $t=0$ with 30\,fs FWHM duration and 30\,J/cm$^2$ fluence. 
While panel a) shows the calculated DoS as used by the simulation and published in \cite{Jain2013,Persson2016nickel}, the colormap in b) shows the occupation of these states over time.
Panel c) shows the number of electrons per atom in the valence band below and above the Fermi level (blue solid and dashed curves, respectively) as well as the average number of core holes and the number of free electrons over time.
Even though the direct interaction with the photons creates core holes via resonant absorption and free electrons via non-resonant absorption, the excitation energy of both processes is so quickly transferred to the valence electrons that only the valence electron distribution ever deviates strongly from the ground state. 
By the end of the pulse in this example, more than half of the $3d$ valence electrons are excited to valence states above the Fermi level, while the highest instantaneous number of core holes was only about one per 100 atoms, as shown in Figure \ref{fig: dos development} c).
Due to the small-bandwidth excitation, the core- and resonant valence states operate like a two-level system. Since the number of resonant valence states is small in comparison to the number of core electrons, the resonant absorption process saturates due to occupied valence states long before the core level is depleted. 
A heated Fermi-Dirac distribution further contributes to the occupation of states above the Fermi level.

% Spectral effects
Since in our experiment, the same monochromatic pulse excites and probes the sample, the situation is different for energies below the edge: absorption only rises after non-resonant absorption has led to sufficient electronic heating until the tail of the hot hole distribution reaches the probed energy.
Only then, additional resonant absorption begins to occur and accelerates further electronic heating and in turn additional pre-edge absorption. Since this process occurs exponentially faster near the absorption edge, it contributes significantly to the observed spectral red-shift (see Figure \ref{fig: effect comparison} a) and b)).

Another cause of the observed edge shift is the shift of the chemical potential $\mu$, which strongly depends on the exact shape of the DoS and is shown in Figure \ref{fig: dos development} b) as a green line.  Initially, $\mu$ increases with absorbed fluence, as thermally excited electrons from the $3d$ states must spread out in energy to the lower DoS above the Fermi level. 
With rising electronic temperature, the high DoS of the $3d$ states becomes less relevant and the chemical potential drops again as expected in regular metals. 
A similar evolution of the chemical potential and electronic temperature  was predicted for optically excited nickel by previous experiments and calculations \cite{bevillon2014free, lin2008electron, lin2007temperature, Humphries2020WDNi}.

% Disagreements to the model, peak absorption deviation
A significant deviation between model and experiment can be observed at the resonance peak itself, where the simulated electron dynamics lead us to expect a much stronger saturation effect than observed experimentally (Figure \ref{fig: effect comparison} c)). 
This underestimation may be related to a fluence-dependent decrease of the excited state lifetime and consequent energetic broadening of the resonant core-valence transition, which is not considered in our model.
While it is unsurprising to find additional resonant effects in the resonance peak itself, the lack of any significant saturation around 852\,eV (Figure \ref{fig: effect comparison} d)) is even more surprising.
Both disagreements point to additional physical effects and call for more sophisticated models.

% Other effects?
We speculatively propose mechanisms which could contribute to these disagreements:
The transition matrix elements could get modified at higher excitation densities, especially around the resonance, while we model the absorption only based on the ground-state spectrum. 
An energy dependence of the electron-electron scattering cross-section could allow for particularly fast scattering of electrons with certain energies, counteracting the saturation. 
Furthermore, a collective, correlated response of the electronic system could modify the DoS or the transitions even on the fast time scale of the FEL pulse duration \cite{Lojewski2022XAS}. Despite these remaining discrepancies, the main aspects of the spectral changes are covered in our very simple population dynamics model.

% Simple explanation for shift
We want to point out that substantially smaller spectral red-shifts were observed before in nickel after excitation with optical lasers, albeit at three orders of magnitude lower excitation fluence. These required qualitatively different interpretations \cite{stamm2007femtosecond, durr2008ultrafast, chang2021electron, Lojewski2022XAS}, where the explanation for time-dependent changes included a variable DoS, calculated using (Time-Dependent) Density Functional Theory (TD)DFT; this dependency is overshadowed in our high-fluence study by the effects of electron population dynamics.

%\section{\label{sec: conclusion} Conclusion}
% What
To summarize, we interpret the fluence-dependent near-edge X-ray absorption spectra of the nickel 2$p_{3/2}$ core level at X-ray fluences of up to 60\,$\mathrm{J/cm^2}$. We propose a rate-equation model, describing the various excitation and decay processes that connect core- and valence electronic states using differential equations based on scaling of known ground-state properties with the evolving electron populations.
% Agreement 
For the measured spectra of metallic nickel, the model successfully predicts the increase of absorption before and its decrease beyond the resonance, as well as the observed shift of the absorption peak over more than three orders of magnitude in fluence.

It therefore allows us to identify the most important processes responsible for spectral changes:
Heating of valence electrons due to secondary electron cascades from Auger electrons, as well as electrons emitted from the valence band due to non-resonant absorption, appeared particularly relevant. Furthermore, saturation appears dominated to by the heated valence states rather than the core holes. 

This study provides the fingerprints of how strong X-ray fluences may alter the electronic system and thus the spectra in studies, where the X-ray pulses were originally assumed to be non-disturbing.
It becomes clear that a complete modeling of high-fluence spectra needs to build upon dominant population dynamics and requires special treatment around resonances. This provides an excellent benchmark for sophisticated theories. 
Our results also apply to the resonant regime which is particularly interesting for pioneering non-linear X-ray studies. 

\begin{acknowledgments}
We acknowledge European XFEL in Schenefeld, Germany, for provision of X-ray free-electron laser beamtime at the SCS instrument and would like to thank the staff for their assistance.
Funding by the Deutsche Forschungsgemeinschaft (DFG, German
Research Foundation) - Project-ID 278162697 - SFB 1242 and the Helmholtz Association (grant VH-NG-1105) is gratefully acknowledged. Access to Synchrotron SOLEIL through proposal ID 20160880 for characterization of static properties of the Ni films is acknowledged.
Parts of this work were funded by the Swiss National Science Foundation (Grants No. PZ00P2-179944)

\section*{Author Contributions}
M.B., C.D., F.D., L.L.G., J.L., J.P.M., B.R. and S.T. conceptualized and planned the experiment;
M.C., C.D., F.D., N.G., L.L.G., M.I., E.J., A.K., C.-H.L., L.M., B.P., B.R., A.S., K.S., C.S., H.W. and A.Y. prepared the measurement apparatus and samples;
O.A., K.A., M.B., J.B., R.C., M.C., V.C., G.S.C., C.D., F.D., R.Y.E., A.E., N.G., L.L.G., O.S.H., M.I., L.M., G.M., P.S.M., B.R., N.R., A.S., J.S., S.T. A.Y. and Z.Y. performed the experiment;
O.A., R.Y.E, L.L.G., O.S.H., B.R. and N.R. analyzed and visualized the results;
M.B. and R.Y.E. wrote the manuscript;
M.B., U.B., C.D., R.Y.E., A.E., L.L.G., O.S.H., M.I., E.J., T.L., P.S.M., K.O., K.R., M.S., J.O.S., S.M.V., H.W. and Z.Y. reviewed and edited the manuscript;
M.B. and J.P.M. supervised or administered the project.
\end{acknowledgments}

\bibliography{bib-nl-xas}% Produces the bibliography via BibTeX.

\end{document}

%\preprint{APS/123-QED}% What does this line do?

%\title{Non-linear Absorption Spectroscopy at the Nickel $L_3$ Edge: Experiment and Modelling}
%\title{Non-linearity in Resonant X-ray Absorption Spectra from Focused FEL Pulses is Dominated by Electron Population Dynamics}
\title{Supplementary Information for: Electron Population Dynamics Dominate Non-linearities in X-ray Absorption Near-Edge Spectra from Focused FEL Pulses}

%\thanks{A footnote to the article title}%

%\keywords{Suggested keywords}%Use showkeys class option if keyword
                              %display desired
\maketitle

%\section{\label{app: shot sorting} Data Analysis}

\section{Data Acquisition}
The experimental X-ray absorption spectra presented here were collected in the scope of a community proposal as the first user-beamtime at the SCS instrument.
% General procedure
The intensities of the two beams generated by the beam-splitting zone plate were recorded using a FastCCD detector \cite{januschek2016performance} with 1920$\times$960 pixels of 30$\times$30\,$\mathrm{\mu m}^2$. 
The intensities of both beams were integrated over a Region Of Interest (ROI) corresponding to 350$\times$350 pixels each to retrieve the signal and reference intensities for each FEL shot.
%, the two beams then impinged on a FastCCD detector with dynamic gain switching
The high beam divergence due to the zone plate focusing distributed the signal on a 4\,mm wide square on the detector 1\,m downstream of the sample, thus greatly decreasing the fluence incident per detector area in order to avoid detector saturation. 
% Low intensity spectra
The gas-attenuator was filled with varying low pressure of nitrogen gas to regulate the transmission through the beamline to the required fluence. We refer to low-intensity spectra if the fluence was consistently below the sample damage threshold and the full measurement could be recorded on a single spot, without scanning the sample.

% High intensity spectra
For measuring high-intensity spectra, the fluence often exceeded the material damage threshold, creating shot craters and sometimes causing larger fractures in the support membrane. For measurements at these fluences, the sample holder was scanned at a speed of 0.5\,$\mathrm{mm\,s}^{-1}$.  Therefore, only about 50\,\% of all FEL shots were transmitted through the windows; in the other cases, one or both beams were blocked or clipped by the frame.
The membranes were arranged on the frame in a periodic pattern of two rows of sample and two rows of reference membranes with a distance of 1\,mm between rows. This ensured that the two FEL foci always impinged on one row of sample and one reference membranes; a third row was unused in between. Therefore, every time the currently scanned rows were incremented, the upper beam would switch from probing reference membranes to probing sample membranes or vice versa, while the opposite holds for the lower beam. 
% Al filter
To prevent detector saturation, an additional aluminum filter of about 13\,$\mu m$ thickness was installed in front of the detector during these measurements. %This filter was not perfectly flat and imparted a notable pattern onto the detector image.

\section{Detector Calibration}
% Detector cooling problem
The temperature of the FastCCD rose consistently during operation and required cool-down periods in between measurements, leading to the temperature varying between -27\,$^\circ$C and -5\,$^\circ$C not only over time during the measurement, but also spatially over the detector area. 
The detector dark signal, as well as the gain coefficients for the three gain settings between which the detector pixels switch automatically, depend on the detector temperature. This made it necessary to reconstruct a temperature-dependent gain calibration. The three temperature-dependent background levels were drawn from dark images collected at various temperatures for each gain setting;  the gain coefficients for each setting were drawn from a statistical analysis of the observed gain switching thresholds, such that the calibrated histogram of pixel intensities becomes continuous over all three gain levels.
While the calibration accounts for the temperature measured using a temperature sensor on the detector, spatial variations over the detector area remain. The primary effect of this temperature variation was a varying background signal, following a spatial exponential distribution between the detector center and rim, with a higher baseline near the detector center. To account for this, an estimated background signal was derived from the measurements themselves: For a running average of 100 images, the illuminated area was cut out and interpolated using fits to the background level in the non-illuminated area. This additional background variation was then integrated into the gain calibration described above.
Furthermore, a mask of hot and dark pixels with irregular behavior was generated from separate measurements and the respective pixels were excluded from the analysis. 
Despite these corrections, the detector inhomogeneities constitute a significant part of measurement uncertainty in the presented spectra. In particular, the uneven warming of the two detector halves on which the upper and lower beam respectively impinged has the potential of introducing systematic uncertainties. Thus, the shot-sorting algorithm described below was applied separately for all rows where the sample was in the upper beam and the reference in the lower and then to all rows where the orientation was reversed. 
Fortunately, differences between the temperature of both detector hemispheres affect the two equally sized groups of data (sample up and sample down) with equal and opposite magnitude.
Therefore, possible systematic deviations are eliminated in the average over both groups and instead contribute to the statistical uncertainty which is represented in the error shown in Figure 2 of the manuscript.
%\ref{fig: spectra} 

\section{Event Classification}
% Shot sorting motivation
Furthermore, whenever the sample or reference membrane was torn due to a particularly intense shot, subsequent shots sometimes impinged on the torn rim of the sample, possibly at an angle to the membrane surface, or did not hit any sample material at all. 
Shots affected in this way were not always trivial to identify from any single measurement parameter, which lead to the following procedure to identify and exclude faulty FEL shots:
First, the detector image in the ROI around one beam was compared to an extended ROI around the other beam using a normalized two-dimensional cross-correlation algorithm. For this, the images were first smoothed by convolution with a Gaussian kernel to remove the influence of the rough surface structure of the aluminum filter before applying the cross-correlation function, both algorithms implemented in the scikit-image \cite{van2014scikit} package. This procedure yields a correlation coefficient and a displacement vector. The correlation coefficient was used as an indicator that both beams were transmitted through a window without significant differences in the wavefront.

% Y-reconstruction
Since data-acquisition of the motor encoder for the position transverse to the scanning direction was dysfunctional, the real path of the beam over the sample frame was reconstructed by combining knowledge of the beam position along the scanning direction and the manual notes in the laboratory book with the correlation coefficients between both beams on the detector. In the later analysis of the damaged samples, this allowed associating specific shot craters scrutinized with SEM-microscopy to specific FEL shots. 

\begin{figure}[ht]
\centering
\includegraphics[width=\linewidth]{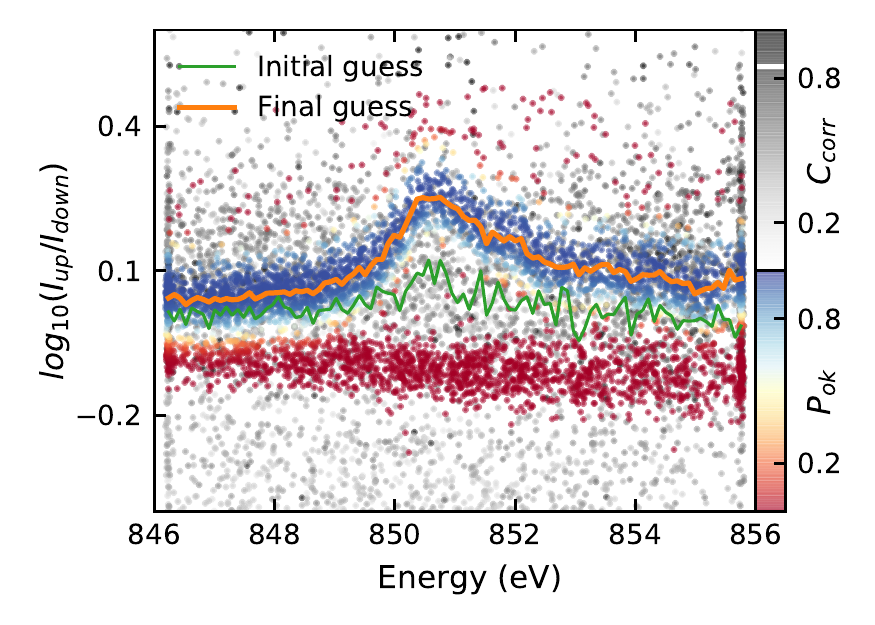}
\caption{
	\textbf{Sorting of FEL-shots}. Gray points represent events which were excluded based on rigid criteria, mainly their correlation coefficient $C_\mathrm{{corr}}$.
	Colored points were analyzed using the iterative GMM optimization. The colorbar shows the estimated posterior probability $P_\mathrm{{ok}}$ that a given shot cleanly probed an unperturbed sample. The green and orange solid lines represent the initial and final average spectrum estimated by the GMM, respectively.}
\label{fig: supp shot sorting}
\end{figure}

% Outliers
From the FEL shots which hit sample windows according to this reconstruction, outliers were dropped if either the correlation coefficient dropped below 85\,\% or the displacement vector deviated by a significant margin (manually calibrated for each measurement setting) from the expected beam-splitting. Likewise, extreme outliers in the ratio of reference to sample intensity were also dropped at this stage. 
% summary
These criteria proved to be largely redundant as they mostly agree with each other on which events to exclude. Using all of these criteria, events where either or both beams are blocked or clipped can be excluded reliably. 
In {Figure \ref{fig: supp shot sorting}}, the thus excluded points are shown in gray.
Each dot represents one FEL shot; its y-position is the logarithmic ratio between sample and reference intensity; the x-position is the photon energy setting of the monochromator, while the gray-scale encodes the Pearson correlation coefficient $\mathrm{C_{corr}}$ of the two regions of interest. 

% GMM
From the distribution of the remaining events, shown in color, it is obvious that further classification is needed.
This is because shots onto a damaged or missing membrane can produce spots with good correlation and the expected displacement vector. These can only be distinguished by the fact that the apparent optical density deviates unreasonably from the expected value at the given photon energy.
However, this transmission ratio is also the quantity of interest for the final spectrum. To disentangle events affected by prior sample damage from true measurements, an iterative approach using a Bayesian Gaussian mixture model was utilized:
To start, an initial guess for a spectrum is computed using the events that passed filter conditions described above and shown as a green line in Figure \ref{fig: supp shot sorting}.
Then, for each FEL shot, the deviation of the logarithmic ratio between sample and reference intensity from the initially guessed spectrum is computed. This allows for generating a histogram of these absorbance deviations from the initial guess. Assuming a good guess of the initial spectrum, one may expect that the ``good" FEL shots are normally distributed around zero deviation, while shots affected by various sources of uncertainty are distributed with some other distribution, dependent on the type of uncertainty. Thus, the histogram is fitted with four Gaussian distributions which are used as prior probability distributions, the first of which corresponds to  ``good" FEL shots. 
Then, the posterior probability of belonging to the category of ``good" shots is computed for each FEL shot. 
These posterior probabilities are used as statistical weights to compute an improved guess of the measured spectrum. 
This improved guess was convoluted with a Gaussian kernel to prevent over-fitting before using it as a new initial guess for the Gaussian mixture model. The spectrum is considered converged when the average change per iteration anywhere in the spectrum is less than 5$\mathrm{\mu OD}$.
This procedure converges (if at all) within a few (between 3 and 20) iterations to a solution that is robust even against a strong variation of the initial guess. In Figure \ref{fig: supp shot sorting}, the thus analyzed shots are shown in colors encoding the final posterior probability $\mathrm{P_{ok}}$, indicating the estimate of the model about the validity of each shot. The final guess of the average spectrum is shown as an orange line.

This final procedure rejects outliers purely based on their deviation from the expected value, which requires further justification: In this case, the procedure should return valid results under the condition that all valid FEL shots at a certain photon energy measure a transmission value within a single continuous range. The width of the distribution is resulting from a combination of non-linear changes and measurement noise. 
This condition must be fulfilled in the present case as long as the fluence-dependent transmission curve of the sample is continuous and a continuous range of pulse energies is contained in the data set (which is the case for SASE fluctuations).
Apart from this logic, the rejected outliers do not appear systematic, further supporting the applicability of the algorithm.

\section{\label{app: spot size} Determination of the effective spot size}
The zone plate has a size of (1\,$\mathrm{mm)^2}$ and combines a focusing Fresnel zone plate,  off-axis by 0.55\,mm with 250\,mm focal length (at 860\,eV) with a line grating with 379.4\,$\mathrm{nm}$ pitch in a single optical element. Details on the zone plate can be found in \cite{Doring2020zoneplates}.
While the monochromator was scanned between 846\,eV and 856\,eV, the effective size of the foci on the sample changed due to the wavelength dependency of the zone plate diffraction.
The photon energy-dependent focal size of the zone plate foci was calculated by ray optics calculations based on the beamline settings \cite{loicsRayOptics}. 
Uncertainties in the exact beam path parameters along the beamline were accounted for by matching the ray optics calculations to effective spot size estimates derived from analyzing the shot craters on the used samples using a simplified form of the procedure laid out in \cite{chalupsky2010spot, sobierajski2013experimental}.
Since a full intensity profile could not be measured from the shot craters, the concept of the effective area of the focal size is used. This area connects the peak fluence $F_0$ with the overall pulse energy $E_\mathrm{{pulse}}$, i.e.
\begin{equation}
    A_{eff} = \frac{E_\mathrm{{pulse}}}{F_0},
    \label{eq: supp effective area}
\end{equation} 
and it can be defined for an arbitrary spot profile.

\begin{figure}[ht!]
    \centering
    \includegraphics{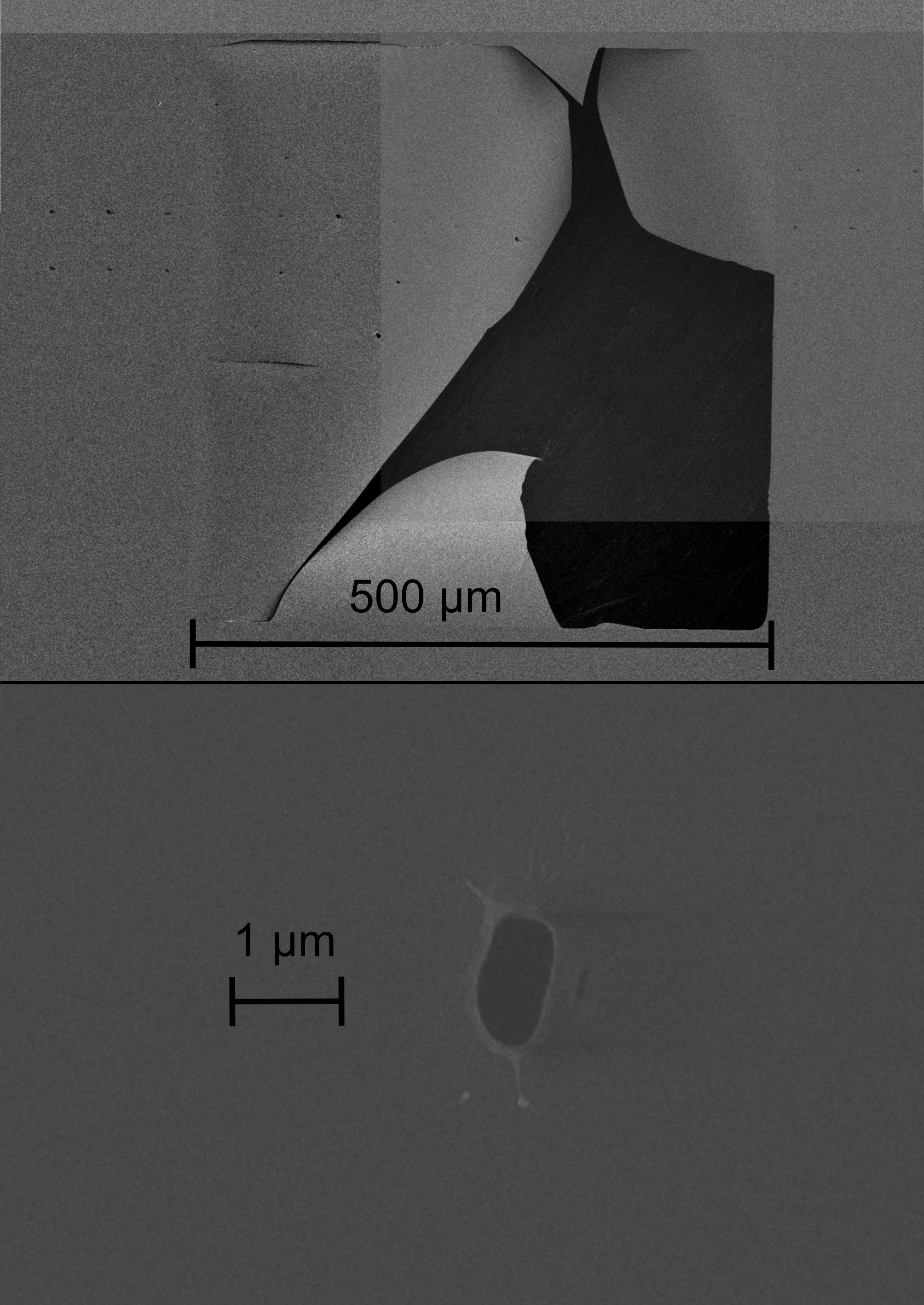}
    \caption{\textbf{SEM images of the used samples} The top image shows a stitched overview image of a nickel film window. One can see rows of FEL imprints as well as the tearing of the membrane. The bottom image shows a single FEL imprint in a SiN reference membrane.}
    \label{fig: SEM}
\end{figure}

To characterize the effective focal size, the reference membranes were analyzed using scanning electron microscopy (SEM). Figure \ref{fig: SEM} shows two SEM images: one overview image of an entire membrane one example of a high-resolution image of a single imprint. Such high-resolution images were taken of 85 selected spots which were associated with the corresponding FEL photon diagnostics data by matching the reconstructed movement of the sample stage to the pattern of imprints on the sample. The reference membranes were chosen for the imprint analysis since their X-ray absorbance can be considered constant for the scanned photon energies.% 

Since the zone plate is a diffractive element, its properties such as focal length are energy-dependent. Furthermore, the fluence-dependent spectra presented in this paper are combined from measurements at two distinct object distances of 250.84\,mm and 252.14\,mm.

\begin{figure*}[ht!]
\centering
\includegraphics[width=\textwidth]{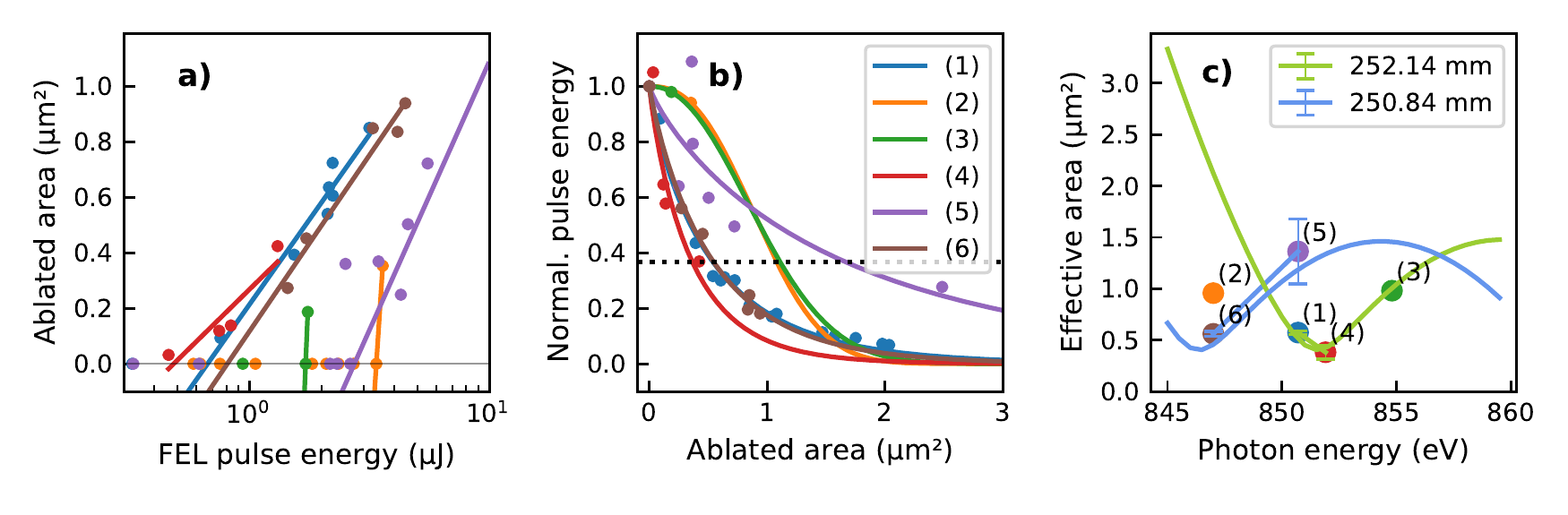}
\caption{
    \textbf{Spot size characterization}. a) Liu's plot to determine the pulse energy damage thresholds. b) Normalized pulse energy plot to determine the effective area. The legend differentiates the spot groups corresponding to table \ref{tab: supp spot size} with colors valid for all three panels. c) Area of FEL foci, comparing the effective area measurements (dots and error bars) to the ray tracing results (line plots) for two focal distances. For groups (2) and (3) no error estimate could be calculated (see text).}
\label{fig: supp spot size}
\end{figure*}

\begin{table*}[ht!]
    \centering
    \begin{scriptsize}
    \begin{tabular}{|c|c|c|c|c|}
    \hline
    Group   &  Object distance (mm)& Photon energy (eV) & Pulse energy threshold ($\mathrm{\mu J}$) & Effective area ($\mathrm{\mu m^2}$)\\
    \hline
    1 & 252.14 & 850.7 & 0.308 & 0.584 \\
    2 & 252.14 & 847.0 & 3.374 & \textcolor{red}{0.957} \\
    3 & 252.14 & 854.8 & 1.712 & \textcolor{red}{0.981} \\
    4 & 252.14 & 851.9 & 0.482 & 0.383 \\
    5 & 250.84 & 850.7 & 2.733 & 1.618 \\
    6 & 250.84 & 847.0 & 0.807 & 0.567 \\
    \hline
    \end{tabular}
    \end{scriptsize}
    \caption{\textbf{Results of the spot size characterization}. The object distance refers to the distance between sample and zone plate. Pulse energies are shown as measured at the XGM without accounting for the efficiency of the zone plate (about 9\,\%) and the transmission of the beamline KB-mirrors (about 80\,\%). The effective area of groups 2 and 3 is under-determined (see text).}
    \label{tab: supp spot size}
\end{table*}

Thus, the SEM images were grouped into six groups (see table \ref{tab: supp spot size}) by focal length and photon energy of the associated shot, and the total damaged or ablated surface area was determined for each shot.
For each group, the minimum FEL pulse energy at which damage is observed on the reference membranes was determined by a Liu's plot \cite{liu1982simple} as shown in Figure \ref{fig: supp spot size}: The ablated area was plotted over the logarithm of the shot energy and a linear fit was applied to the shots with less than 1\,$\mathrm{\mu m}^2$ ablated area to determine the pulse energy damage threshold.\\
Following the concept outlined in \cite{chalupsky2010spot}, the pulse energy of all shots was then normalized using this damage threshold to derive the normalized fluence level $f(S)$ as a function of the ablated area $S$. If the spot size were Gaussian, the function $f(S)$ should be a simple negative exponential, and the effective area would correspond to the value of $S$ where $f(S)$ equals $1/e$, which is indicated in Figure \ref{fig: supp spot size} b) as a horizontal dotted line.
For the given non-Gaussian case, the function $f(S)$ was fitted with a modified exponential $f(S) = e^{(-a S^b)}$, yielding for each group of shots fit parameters $a$ and $b$ and their unceratinties $\sigma_a$ and $\sigma_b$.
The effective area is then calculated as the integral of $f(S)$. For the estimate of the uncertainty shown as error bars of Figure \ref{fig: supp spot size}  c), the integral was also evaluated for $a + \sigma_a$ and $b+\sigma_b$ as well as $a - \sigma_a$ and $b-\sigma_b$. Groups (2) and (3) each contained only a single FEL shot with visible damage. While this does allow for an estimate of the damage threshold (compare Figure \ref{fig: supp spot size} a)), the function $f(S)$ shown in b) is barely determined by fitting only to the normalization point and a single further point. Thus, no mathematical uncertainty estimate can be given and resulting points (2) and (3) shown without error bars in Figure \ref{fig: supp spot size} c) should be considered as tentative estimates.\\
The ray optics calculation tracks the beams in the vertical and horizontal planes. The focus area was calculated from the ray optics calculations assuming an elliptical beam shape. It revealed a slightly astigmatic focus on the sample, due to the beamline optics using separate horizontal and vertical focusing, thus illuminating the zone plate with slightly non-uniform beam divergence. The separate location of horizontal and vertical foci leads to the two minima visible in Figure \ref{fig: supp spot size}. Based on the measurements, a minimal beam-waist radius of 80\,nm was imposed on the ray tracing calculations, accounting for imperfections in beam quality and optics.\\
The fluences for the high-fluence spectra shown in this paper are calculated based on the pulse energy measured by the XGM behind the monochromator, multiplied with the efficiency of the focusing optics (80\,\%) \cite{Mercurio2022real} and zone plate (9\,\%) and divided by the effective area derived with the presented ray optics calculations. 
Since the XGM was calibrated for high pulse energies, the fluences for the low-fluence spectra are based on the reference intensities measured on the CCD and calibrated to be consistent with the reference intensities measured for the high-fluence data, accounting for the calculated transmission of the aluminum filter.

\bibliography{bib_supplement}% Produces the bibliography via BibTeX.